\documentclass[aps,twocolumn]{revtex4-1}
\usepackage{amsmath, amssymb, amsbsy, amsfonts}
\usepackage{graphicx}
\usepackage{physics}
\usepackage[colorlinks,linkcolor=blue,citecolor=blue,urlcolor=blue]{hyperref}

\begin{document}
\title{An eternal discrete time crystal beating the Heisenberg limit}
\author{Changyuan Lyu$^1$}
\author{Sayan Choudhury$^1$}
\author{Chenwei Lv$^1$}
\author{Yangqian Yan$^{1,2}$}
\email{yan281@purdue.edu}
\author{Qi Zhou$^{1,2,3}$}
\email{zhou753@purdue.edu}
\affiliation{1. Department of Physics and Astronomy, Purdue University, 525 Northwestern Avenue, West Lafayette, IN 47907, USA\\
2. Center for Science of Information, Purdue University, West Lafayette, IN 47907, USA\\
3. Purdue Quantum Science and Engineering Institute, Purdue University, 1205 W State St, 
West Lafayette, West Lafayette, IN 47907, USA
}
\date{\today}

\begin{abstract}
  A discrete time crystal (DTC)  repeats itself with a rigid rhythm, mimicking a
  ticking clock set by the interplay between its internal structures and an 
  external force \cite{Wilczek2012, ZhangX2012, Oshikawa2015, Sondhi2016PRL, Nayak2016, NormanYao2017,Lukin2017,Monroe2017, Abanin2017, Sreejith2018, Sean2018PRL, Huang2018}. DTCs promise profound 
  applications in precision time-keeping and other quantum techniques. However, 
  it has been facing a grand challenge of thermalization.  The periodic driving 
  supplying the power may ultimately bring DTCs to thermal equilibrium and destroy
  their coherence \cite{Rigol2014, Ponte2015, Moessner2014}. Here, we show that 
  an all-to-all interaction delivers a 
  DTC that evades thermalization and maintains quantum coherence and quantum 
  synchronization regardless of spatial inhomogeneities in the driving field and 
  the environment. Moreover, the sensitivity of this DTC scales with the total 
  particle number to the power of three over two, realizing a quantum device of 
  measuring the driving frequency or the interaction strength beyond the Heisenberg 
  limit. Our work paves the way for designing novel non-equilibrium phases with 
  long coherence time to advance quantum metrology. 
 \end{abstract}

 \maketitle

A periodic driving may continuously pump energies into a DTC and eventually heat 
it up to the infinite temperature \cite{Rigol2014, Ponte2015, Moessner2014}. 
A number of schemes have been proposed to slow down the thermalization
\cite{Sondhi2016PRL, Nayak2016, Abanin2017, NormanYao2017}, 
such as the many-body localization (MBL), the Floquet prethermalization and 
crypto-equilibrium. Compared with other schemes only retaining the coherence of 
DTCs within certain time scales, MBL is of particular interest. Disorder breaks 
an interacting system into localized l-bits to encode the memory of the initial 
state \cite{Huse2010}, and suppresses thermalization up to an arbitrarily long time scale. 
However, most studies have considered homogeneous drivings so far. In practice, 
the driving field may vary across a DTC and local perturbations may further 
amplify the spatial inhomogeneities, both preventing individual constituents of 
the DTC from synchronization and impeding applying DTCs in quantum technologies. 
Whereas MBL could stabilize a DTC against weak inhomogeneous perturbations
to $\pi$-rotations \cite{Sondhi2016PRB}, it is no longer powerful in the presence of strong inhomogeneities, as the 
exponentially decayed couplings between l-bits in MBL have readily weakened the 
synchronization between remote parts of a DTC in spite of the presence of interactions.

Fundamental questions naturally arise. (1) How to access a DTC that could maintain 
quantum coherence and quantum synchronization in the presence of arbitrarily strong
inhomogeneous driving fields and local perturbations? (2) Furthermore, how to 
implement such a DTC to  promote the precision of quantum metrology? 
 
\begin{figure} 
  \includegraphics[width=0.48\textwidth]{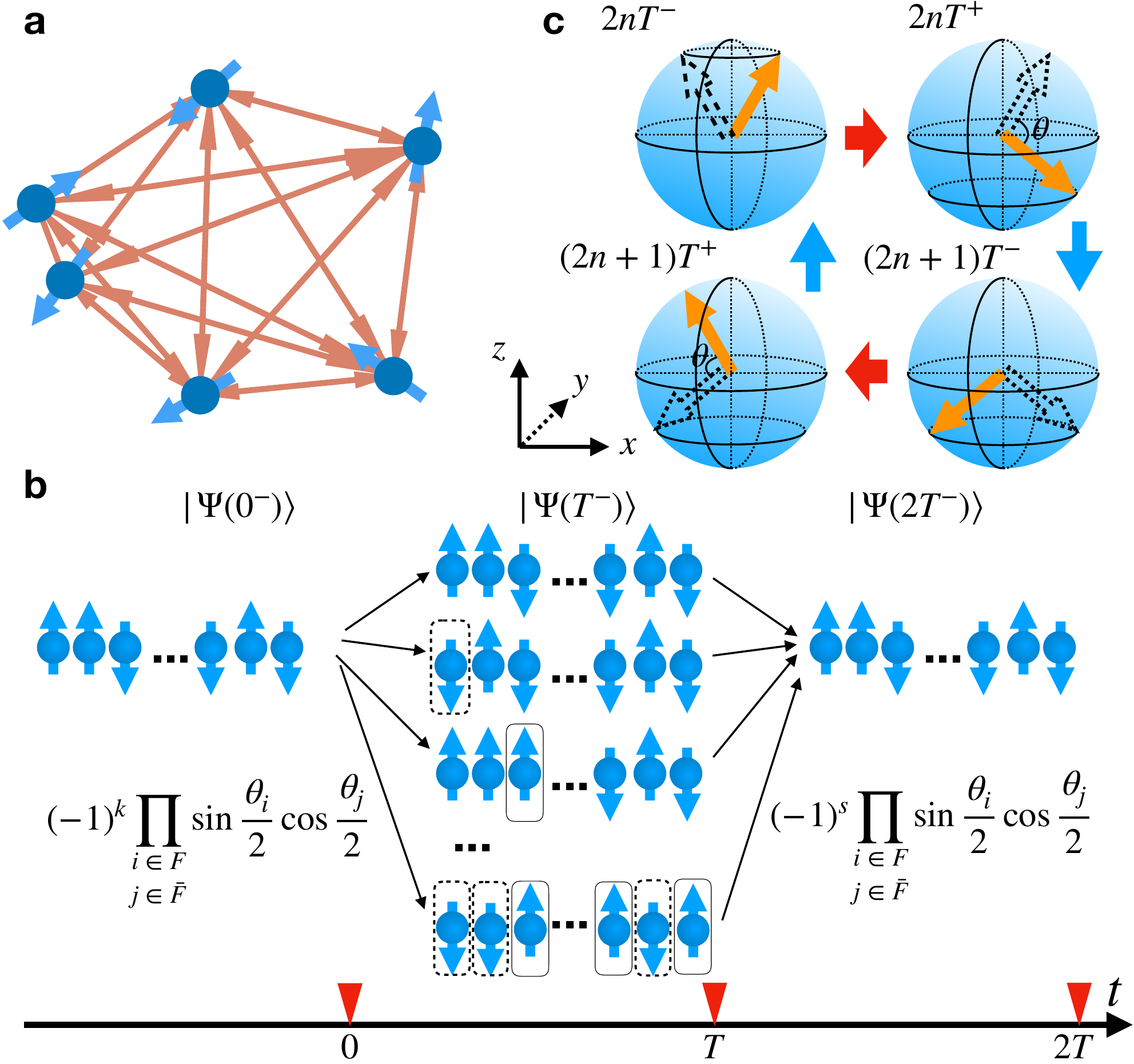}
  \caption{\label{fig1}A DTC induced by an all-to-all interaction. (a) Arrows 
  represent an all-to-all interaction between spin-$1/2$s (blue spheres attached 
  to arrows). (b) A perfect revival of an arbitrary initial state due to the 
  constructive interference among all pathways. Dashed and solid boxes highlight 
  the $k$ spin-ups and $s$ spin-downs flipped by the first pulse, which lead to 
  the geometric phase,  $(-1)^k$ and $(-1)^s$, respectively. (c) Rotations of 
  the spin-$L$ (yellow arrow) on the Bloch sphere. When $JT=\pi$, the non-linear 
  term, $J L_z^2$, leads to an effective $\pi$ rotation about the $z$ axis between 
  $2nT^+$ and $(2n+1)T^-$ such that any initial state returns to the initial 
  position after $2T$ for any $\theta$. Triangles on the time axis represent $H_{\text{pul}}$.}
\end{figure}

We consider $N$ spin-1/2s described by a Hamiltonian, 
$H = H_{\text{int}} + \sum_n H_{\text{pul}} \delta(t-nT)$, 
where
\begin{align}
H_{\text{int}} &= 2J \sum_{i<j} S^{z}_i S^{z}_j,  \label{H}\\
H_{\text{pul}} &= \sum_i^N \theta_{i} {S}^y_i.
\end{align}
As shown in Fig.~\ref{fig1}(a), $J$ is the strength of an all-to-all interaction, 
which has been considered in the Lipkin-Meshkov-Glick model \cite{Lipkin1965}. 
$\vec{S}_i = \frac{1}{2} \vec{\sigma}_i$ and $\vec{\sigma}_i$ are Pauli matrices 
(we have set $\hbar=1$). 
Eq.~(\ref{H}) can be realized using 
spin-1/2s coupled to a cavity or a waveguide \cite{Hung201603777, Esslinger2013}, or 
particles with long-range interactions whose ranges are much larger than the system size. 
The equivalence between spin-1/2s and bosons also provides a natural realization 
of such interaction \cite{Fazio2017}. $\theta_{i}$ determines the angle rotated 
by the $i$th spin about the $y$-axis. The dependence of 
$\theta_{i}$ on $i$  characterizes the spatial inhomogeneity of the rotations. 
 
We prove that, when $JT=\pi$ is satisfied, any initial state returns to itself at 
$t=2nT^-$ for any { even} $N \in 2\mathbb{Z}$ and any $\theta_{i}$ as an arbitrary function 
of $i$. $t^-$ ($t^+$) denotes the time right before (after) a pulse is applied. 
This perfect revival delivers an eternal DTC that evades thermalization and is
equipped with a strong synchronization 
even in the presence of a noisy environment. Previous works on normalized all-to-all 
interactions have considered the small $J$ limit of Eq.~(\ref{H}) 
\cite{Fazio2017},
not the optimal choice of $JT$ discussed here. 

Consider an initial state with $m$ spin-ups and $N-m$ spin-downs, 
$\ket{\Psi(0^-)}=\prod_i\ket{\eta}_i$, where $\eta=\uparrow, \downarrow$.
After the first pulse, 
\begin{align}
\ket{\uparrow}_i&\rightarrow +\cos(\frac{\theta_{i}}{2}) \ket{\uparrow}_i+\sin(\frac{\theta_{i}}{2})\ket{\downarrow}_i,\\
\ket{\downarrow}_i&\rightarrow -\sin(\frac{\theta_{i}}{2}) \ket{\uparrow}_i+\cos(\frac{\theta_{i}}{2}) \ket{\downarrow}_i,
\end{align}
$\ket{\Psi(0^+)}$ becomes a superposition of $2^N$ states, each of which is 
obtained from flipping $s$ spin-ups and $k$ spin-downs of $\ket{\Psi(0^-)}$, 
as shown in Fig.~\ref{fig1}(b). Each state acquires a dynamical phase, 
$e^{-i\varphi_1}$, imposed by $H_{\text{int}}$ from $t=0^+$ to $t=T^-$.  
The second pulse flips the spins again, followed by $H_{\text{int}}$ imposing 
another dynamical phase, $e^{-i\varphi_2}$, from $t=T^+$ to $t=2T^-$, and  
\begin{equation}
\ket{\Psi(2T^-)} = A\ket{\Psi(0^-)}+..., \label{EA}
\end{equation}
where $...$ represents states different from $\ket{\Psi(0^-)}$. 

To return to $\ket{\Psi(0^-)}$, the $s$ ($k$) spin-ups (spin-downs) flipped  by 
the first pulse need to be flipped back to spin-ups (spin-downs) during the second pulse. 
$2^N$ such pathways allow the system to come back to $\ket{\Psi(0^-)}$. 
The contribution to $A$ from each pathway is written as 
$(-1)^{k+s}\prod_{j\in \bar{F}} \cos^2(\frac{\theta_{j}}{2})\prod_{i\in {F}} \sin^2(\frac{\theta_{i}}{2})$, where $(-1)^{k+s}$ comes from flipping $k+s$ spin-1/2s twice, equivalent to the 
geometric phase from rotating these spins about the $y$ axis for $2\pi$. $F$ ($\bar{F}$) 
denotes the collection of flipped (unflipped) spins. As each of these $2^N$ states is an
eigenstate of $H_{\text{int}}$, $\varphi_1=(m-s+k)(m-s+k-N)\pi$, and 
$\varphi_2=m(m-N)\pi$ when $JT=\pi$. $m$-independent terms have been dropped. 
The total dynamical phase accumulated from $0^-$ to $2T^{-}$ is 
$e^{-i(\varphi_1+\varphi_2)}=e^{i\pi\{2[m^2+m(k-s-N)-ks]+N(s-k)+k^2+s^2\}}=(-1)^{k+s}$. 
We have used  $N\in 2\mathbb{Z}$, and $e^{iZ^2\pi}=e^{i Z\pi}=(-1)^Z$ for any integer $Z$. 
This dynamical phase factor cancels exactly the previously obtained geometric phase, 
and thus $A=\sum_{F}\prod_{j\in \bar{F}} \cos^2(\frac{\theta_{j}}{2})\prod_{i\in {F}}\sin^2(\frac{\theta_{i}}{2})$. 
$\sum_{F}$ denotes the sum over all $2^N$ choices of flipping the $N$ spins in 
$\ket{\Psi(0^-)}$. Since $F$ is an arbitrary choice from the $N$ spins,
\begin{align}
A=\prod_i(\sin^2(\frac{\theta_{i}}{2})+\cos^2(\frac{\theta_{i}}{2}))=1.
\end{align}
These discussions apply to any initial product state and any $t\in [2nT^-, 2(n+1)T^-]$.
Thus, any initial state returns to itself at $t=2nT^-$.  
Unlike traditional spin-echo schemes using tailored pulses to restore quantum coherence \cite{Yan2013},
we implement interactions, one source of the decoherence, to overcome the other, 
the inhomogeneities, so as to access a perfect dynamical localization, an analogy 
to the Anderson localization in the Hilbert space \cite{DAlessio2013}. Therefore, this interaction 
induced spin-echo could be used in a broad class of systems to extend the coherence time.  

For spatially uniform pulses, a simpler proof exists. $H$ is rewritten as 
\begin{align}
  H_{\text{hom}}=J L_z^2+\theta L_y\sum_n\delta(t-nT),\label{HH}
\end{align}
where $\vec{L}=\sum_i\vec{S}_i$. Eq.~(\ref{HH}) is equivalent to the kicked top 
model describing a periodically driven spin-$L$ \cite{Haake1987}, 
{ where $L=\frac{N}{2}$}. The propagator from 
$t=2nT^-$ to $t=2(n+1)T^-$ is written as
\begin{equation}
U_{JT}(2T)=e^{-iJTL_z^2}e^{-i\theta L_y}e^{-iJT L_z^2}e^{-i\theta L_y}.
\end{equation} 
As $e^{-i\pi L_z^2}=e^{-i\pi L_z}$ applies to any integer $L$ (or even $N$), 
$U_\pi(2T)=e^{-2i\pi L_z}e^{e^{i\pi L_z}(-i\theta L_y)e^{-i\pi L_z}}e^{-i\theta L_y}=1$.
As shown by Fig.~\ref{fig1}(c), 
any state on the Bloch sphere of a spin-$L$ returns to the original place after 
$2T$. If $N\in 2\mathbb{Z}+1$, $e^{-i\pi L_z^2}$ and $e^{-i\pi L_z}$ are no longer identical, 
and such DTC with a period of $2T$ does not exist. { In contrast, if we consider spin-1 instead of spin-1/2 in Eq.~(\ref{H}), such even-odd effect is absent, as $L$ is always an integer for both even and odd $N$.  }

$U_\pi(2T)=1$ means that the quasi-energy spectrum of $H_{\text{eff}}$, where 
$U_{JT}(2T)=e^{-i H_{\text{eff}}}$, has $2^N$ degenerate eigenstates. Whereas this 
looks similar to the non-interacting case when $\theta_i=\pi$, a conceptual 
difference is that, the degeneracy here is stable against any perturbations in 
$\theta_i$, unlike non-interacting systems, where any infinitesimal derivation 
from a homogeneous $\pi$-pulse lifts the degeneracy,  breaks the integrability, 
and suppresses DTCs. 
 
\begin{figure*} 
  \includegraphics[angle=0,width=\textwidth]{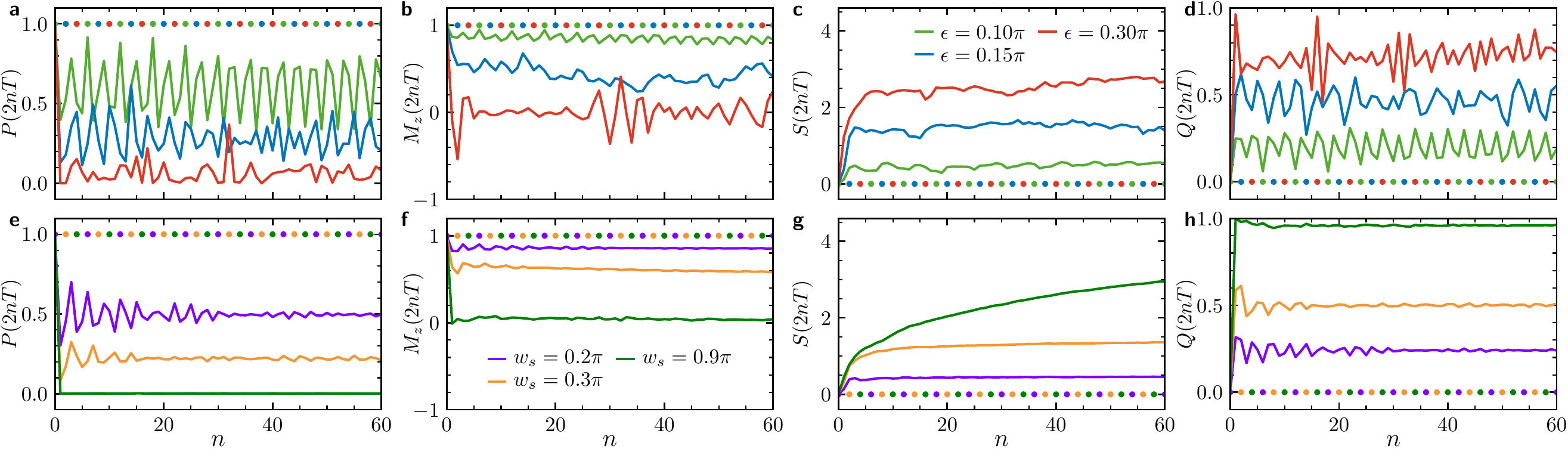}
  \caption{\label{fig2} Comparison between the all-to-all interaction and a power-law 
  potential with $\alpha=3$. Here $N=14$. (a-d) Uniform rotations of spins, $w_s=0$.  The DTC 
  with all-to-all interactions (dots) is unaffected by $\epsilon$, the derivation 
  of $\bar{\theta}$ from $\pi$. With the power-law potential (curves), increasing  
  $\epsilon$ leads to the suppression of $P(2nT)$ and $M_z(2nT)$, and 
  the growth of $S(2nT)$ and $Q(2nT)$. (e-h) Keeping $\bar{\theta}=\pi$ and increasing the spatial
  inhomogeneities $w_s$, the DTC with the power-law potential is 
  suppressed. The DTC with all-to-all interactions remains stable.}
\end{figure*}

To highlight the stability 
against the spatial inhomogeneity, we compare the all-to-all interaction model to
the power-law interaction model, $H'= H_{\text {int}}'+\sum_n H_{\text{pul}} \delta(t-nT)$, where
\begin{align}
H_{\text {int}}'=2J\sum_{i<j}\frac{S_i^zS_j^z}{|i-j|^\alpha}.\label{Hintp}
\end{align}
Starting from $\ket{\Psi(0^-)}=\prod_i\ket{\uparrow}_i$, 
we compute some quantities for both interactions using exact diagonalization, 
\begin{align}
  P(2nT^-)&=|\bra{\Psi(0^-)}\ket{\Psi(2nT^-)}|^2,\\
  M_z(2nT^-)&=2\bra{\Psi(2nT^-)}L_z\ket{\Psi(2nT^-)}/N,\\
  E(2nT^-)&=\bra{\Psi(2nT^-)}H_{\text{int} }\ket{\Psi(2nT^-)},\\
  S(2nT^-)&=-\mathrm{Tr}(\rho_B\ln \rho_B).
\end{align}
$P(2nT^-)$ characterizes the quantum memory of the initial state, $M_z(2nT^-)$ 
denotes the $z$-component of the total spin, 
$E(2nT^-)$ (or $E'(2nT^-)=\bra{\Psi(2nT^-)}H'_{\text{int} }\ket{\Psi(2nT^-)}$) 
captures the absorption of energy, and $S(2nT^-)$ is the bipartite entanglement 
entropy using the reduced density matrix of half of the system, $\rho_{B}$.
 
When $\theta_{i}=\bar\theta$ for any $i$, a finite $J$ in Eq.~(\ref{Hintp}) 
restores the quantum coherence, if $\epsilon=\bar\theta-\pi$ is small
\cite{NormanYao2017,Lukin2017,Monroe2017}. However, with increasing $\epsilon$, 
both $P(2nT^-)$ and $M_z(2nT^-)$ get suppressed, 
as depicted in Fig.~\ref{fig2}(a-d).  Meanwhile, $Q$ and $S$ grow quickly, where
we have used $Q= \frac{E(2nT^-)-E(0)}{E_{\infty}-E(0)}$ to characterize the absorption 
of the energy. $E_{\infty}=2^{-N}\sum_{j}\bra{j}H_{\text{int}}\ket{j} $ is the 
energy at the infinite temperature and $\ket{j}$ denotes the $2^N$ eigenstates of 
$H_{\text{int}}$. These results signify the thermalization at large $\epsilon$. 
We further take into account the spatial
inhomogeneity. As shown in Fig.~\ref{fig2}(e-h), we choose a random $\theta_i$ 
from $[\bar{\theta}-w_s, \bar{\theta}+w_s]$ with a constant probability.
When $w_s$ is finite, the thermalization becomes even faster 
and $Q$ approaches 1, indicating that the system thermalizes to the infinite temperature.
Adding onsite disorder to { introduce} MBL does 
not change qualitative results (Supplementary Material). In contrast, $P(2nT^-)$ 
and $M_z(2nT^-)$ of the all-to-all interaction are unaffected by $w_s$ and remain 
unity, and both $Q(2nT^-)$ and $S(2nT^-)$ remain zero,  
directly reflecting the robustness of this eternal DTC against arbitrarily strong
spatial inhomogeneities and representing the most synchronized DTC.  
 
\begin{figure*} 
  \includegraphics[width=\textwidth]{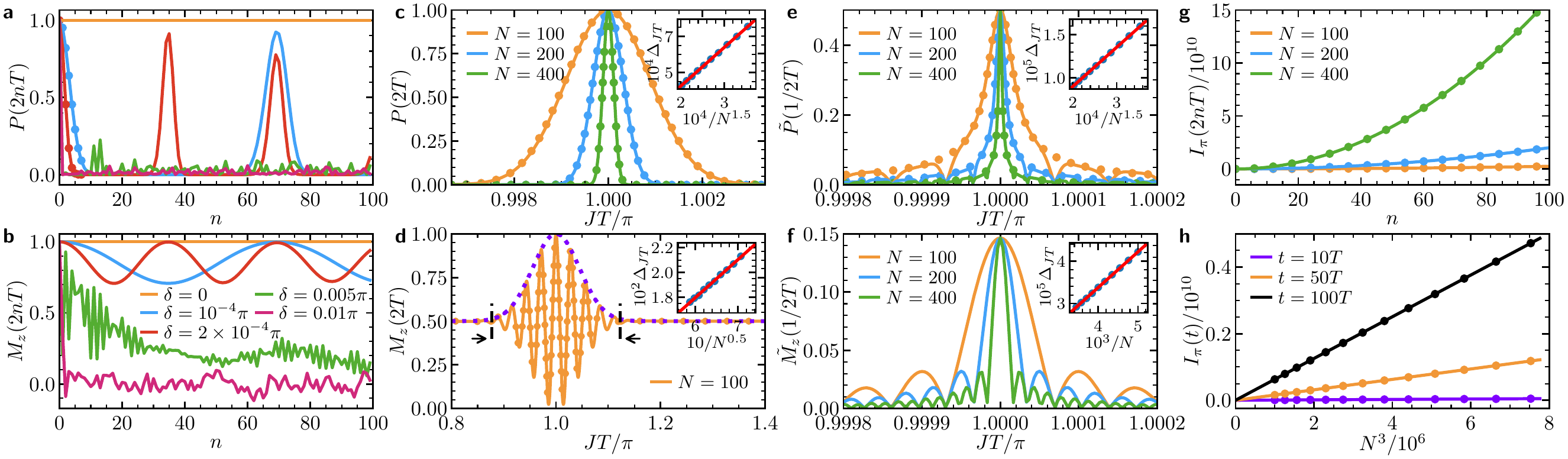}
  \caption{\label{fig3} Sensitivity to $JT$. Curves (dots) are numerical (analytical) results. (a,b) $P(2nT)$ and $M_z(2nT)$ as functions of $n$ 
  at various $JT$.  When $|JT-\pi|\gg \pi/N^{{3}/{2}}$, both quantities quickly decrease down 
  to zero. $N=200$ has been used in the calculation. (c,d)  $P(2T)$ and $M_z(2T)$ at a fixed time $t=2T$ as a function of $JT$. For a fixed $N$, { both $P(2T)$ and $M_z(2T)$ have narrow peaks centered at $JT=\pi$. Whereas $M_z(2T)$ has an additional fast oscillation, the dashed curve highlights the analytical result of its profile, whose width is denoted by black arrows}. Insets show the scalings of the widths of the peaks with $N$. (e,f)
  The power spectra, $\tilde{P}(1/2T)$ and $\tilde{M}_z(1/2T)$, which contain information of both short 
  and long-time dynamics, are also featured by narrow peaks around $JT=\pi$. 
  $M=200$ is used in numerics.
  Insets show the scaling of the widths of the peaks with $N$.
  (g) The quantum Fisher information $I_{\pi}(2nT)$ as a function of $n$.
  (h) $I_\pi(2nT)$ is proportional to $N^3$. $\theta_i=\pi/4$ is used in all panels.} 
\end{figure*}

We now discuss applications of this DTC. As aforementioned, the perfect revival 
at $t=2nT^-$ comes from the same dynamical phase of all  $2^N$ pathways of returning 
to $\ket{\Psi(0)}$ when  $JT= \pi$. Once $JT\neq \pi$, these dynamical phases are 
no longer the same. In particular, the larger $N$ is, the more rapidly the dynamical 
phase varies with changing the pathways. In the large $N$ limit, this DTC becomes 
supersensitive to the value of $JT$ and serves as a high precision device to measure 
either $J$ or $T$. 

Since it is time-consuming to  solve more than $14$ lattice sites using exact 
diagonalization when inhomogeneities exist, we focus on homogeneous systems. It 
is expected that the { lower bound} of the results of an inhomogeneous 
distribution, $\theta_i\in [\bar{\theta}-w_s,\bar{\theta}+w_s]$, could be estimated 
using homogeneous $\theta_i=\bar{\theta}\pm w_s$.  As an example, we consider 
$\theta_{i}$ fixed at $\pi/4$. As shown in Fig.~\ref{fig3}(a), $P(2nT^-)$ 
quickly vanishes if $|\delta|\gg\pi/N^{\frac{3}{2}}$, where $\delta = JT-\pi$. It 
is known that the Heisenberg limit, $1/N$, sets the bound of the precision in 
linear metrology, whereas non-linearity allows going beyond this limit {\cite{Pirandola2018}}. The DTC 
discussed here represents a new category of nonlinear quantum metrology using  
periodic drivings.
  
We evaluate some observables to quantitatively characterize the sensitivity. 
$P(2T)$, the returning probability to $\ket{\Psi(0^-)}$ after two periods, 
captures short time dynamics. Fig.~\ref{fig3}(c) shows that the dependence of $P(2T)$ on 
$JT$ has a narrow peak centered at $\pi$, whose width is of the order of 
$1/N^{\frac{3}{2}}$.  Such scaling can be obtained analytically (Supplementary 
Materials), and is verified numerically, as shown in the inset of Fig.~\ref{fig3}(c). Another quantity is the power spectrum, $\tilde{P}(f)=\frac{1}{M}\sum_{n=0}^{M-1}e^{i2\pi nTf}P(nT)$. 
We are particularly interested in $\tilde{P}(f=\frac{1}{2T})$ 
characterizing the response of the DTC at half of the frequency of the periodic 
driving. The dependence of $\tilde{P}(f)$ on $JT$ also has a peak around $\pi$. 
We define the full width at half maximum as $\Delta_{JT}$, and find both numerically 
and analytically that $\Delta_{JT}$ is proportional to $1/N^{\frac{3}{2}}$ 
(Supplementary Material). 

To gain insights into the scalings, we consider the quantum Fisher information,
\begin{align}
&I_{JT}(2nT)=\lim_{\epsilon\rightarrow 0}4\frac{1-F_{\epsilon}}{\epsilon^2},\\
&F_{\epsilon}=|\bra{\Psi(0^-)}U_{JT}(2nT)U_{JT+\epsilon}(-2nT)\ket{\Psi(0^-)}|^2,
\end{align}
where $F_{\epsilon}$ is the Loschmidt echo. The squared root of the quantum Fisher 
information limits the precision of a phase measurement \cite{Helstrom1969}. The uncertainty of $JT$ 
is bounded by $\sqrt{I_{JT}(2nT)}$, i.e., $\Delta_{JT}\ge 1/\sqrt{I_{JT}(2nT)}$. 
We have found analytically that (Supplementary Materials), 
\begin{equation}
I_{\pi}(2nT)=\frac{n^2}{4}[\sin^2(2\bar{\theta})N^3+ 2\sin^4(\bar{\theta})N^2]. 
\end{equation}
When $\bar{\theta}\neq 0, \pm \pi/2, \pi $, $I_{\pi}(2nT)\sim n^2 N^3$, provided 
that $\sin^2(2\bar{\theta})N^3\gg2\sin^4(\bar{\theta})N^2$. Thus, $I_{\pi}(2nT)$ 
scales with $n^2N^3$ in the large $N$ limit, as shown in Fig.~\ref{fig3}(g,h).
Correspondingly, $\Delta_{JT}\ge 1/\sqrt{I_{\pi}(2nT)}\sim n^{-1}N^{-\frac{3}{2}}$. 
This is precisely what we have obtained in Fig.~\ref{fig3}(c,e).

DTCs previously discussed in the literature are stable within a finite range of 
both the interaction strength and a uniform derivation of $\theta_i$ from $\pi$.  
In contrast, the all-to-all interaction induced DTC is stable against any spatial 
fluctuations in $\theta_i$ and meanwhile supersensitive to $JT$. In practice, it 
is much easier to control $J$ and $T$ other than the $N$ local parameters $\theta_i$ 
in a noisy environment, where $\theta_i$s may not have any correlations at different 
locations. Moreover, our DTC could be used to measure $JT$ with high precision { beyond the Heisenberg limit}. It 
mimics a supersensitive clock. If the frequency of the external field, 
$\omega_{\text{\text{d}}}=1/T$, is fixed,  $J$, which corresponds to some internal 
parameter of a clock, for instance, the length of a pendulum clock, needs to be 
tuned with a precision of $1/N^{\frac{3}{2}}$  
to deliver rigid ticks at $t=2nT$. Otherwise, this DTC stalls to avoid errors in 
the time-keeping. Our results thus lead to a new type of precision measurement of 
$J$.
From $JT=\pi$, the precision of $J$ can be estimated as 
${\Delta_J}/{J}\approx \Delta_{\text{d}}/\omega_{\text{d}}+{N^{-\frac{3}{2}}}$,
where $\Delta_{\text{d}}/\omega_{\text{d}}$ characterizes 
the precision of the driving frequency. When 
${N^{-\frac{3}{2}}}\gg \Delta_{\text{d}}/\omega_{\text{d}}$, ${\Delta_J}/{J}$
scalings with $N^{-\frac{3}{2}}$. When 
$N\rightarrow (\Delta_{\text{d}}/\omega_{\text{d}})^{-2/3}$, 
the uncertainty of $J$ eventually approaches the precision limit of 
$\Delta_{\text{d}}/\omega_{\text{\text{d}}}$. { Whereas the precision of $\Delta_{\text{d}}/\omega_{\text{\text{d}}}$ is up to $10^{-19}$ in the THz regime \cite{YeJun2018}, typical experiments on ultracold atoms, ion traps and NV centers have interaction strengths $\sim 10^{2}-10^{5} $Hz. In such regime, the precision of $\Delta_{\text{d}}/\omega_{\text{\text{d}}}$ could be $10^{-6}$ and above.} Our results thus provide a new application of precision time-keeping in many-body physics.

Alternatively, if $J$ is fixed, the DTC discussed here could gauge the frequency, 
as only a driving field, whose $T$ deviates from $\pi/J$ within $1/N^{\frac{3}{2}}$, 
could induce its long-lasting dynamics. Different from atomic clocks using a 
transition with a narrow line width, the selection of the driving frequency here entirely 
comes from the many-body effect we previously discussed. In particular, the rotated 
angle, $\bar{\theta}$, can be arbitrary such that the DTC could function in a 
non-ideal environment, unlike previous works requiring a precise control of pulses 
in non-linear metrology without periodic driving \cite{Rey2007, Sundaram2008, Napolitano2011}. Though $1/J$ may not be as 
precise as transition frequencies in atomic clocks, the many-body effect induced 
$1/N^{\frac{3}{2}}$ 
scaling could make this DTC a useful gauge of the frequency or time.

We have also studied the scalings of other quantities. We have found that $M_z(2T^-)$ 
and $\tilde{M}(\frac{1}{2T})$ scale with $1/N^{\frac{1}{2}}$ and $1/N$, respectively,  
as shown by Fig.~\ref{fig3}(b,d,f). Similar scalings are obtained for other 
uniform rotations. For instance, when $\theta_i=\pi/2$, $\Delta_{JT}$
of either $\tilde{P}(\frac{1}{2T})$ 
or $\tilde{M}_z(\frac{1}{2T})$ scales with $1/N$ (Supplementary Materials).

\begin{figure} 
  \includegraphics[width=0.483\textwidth]{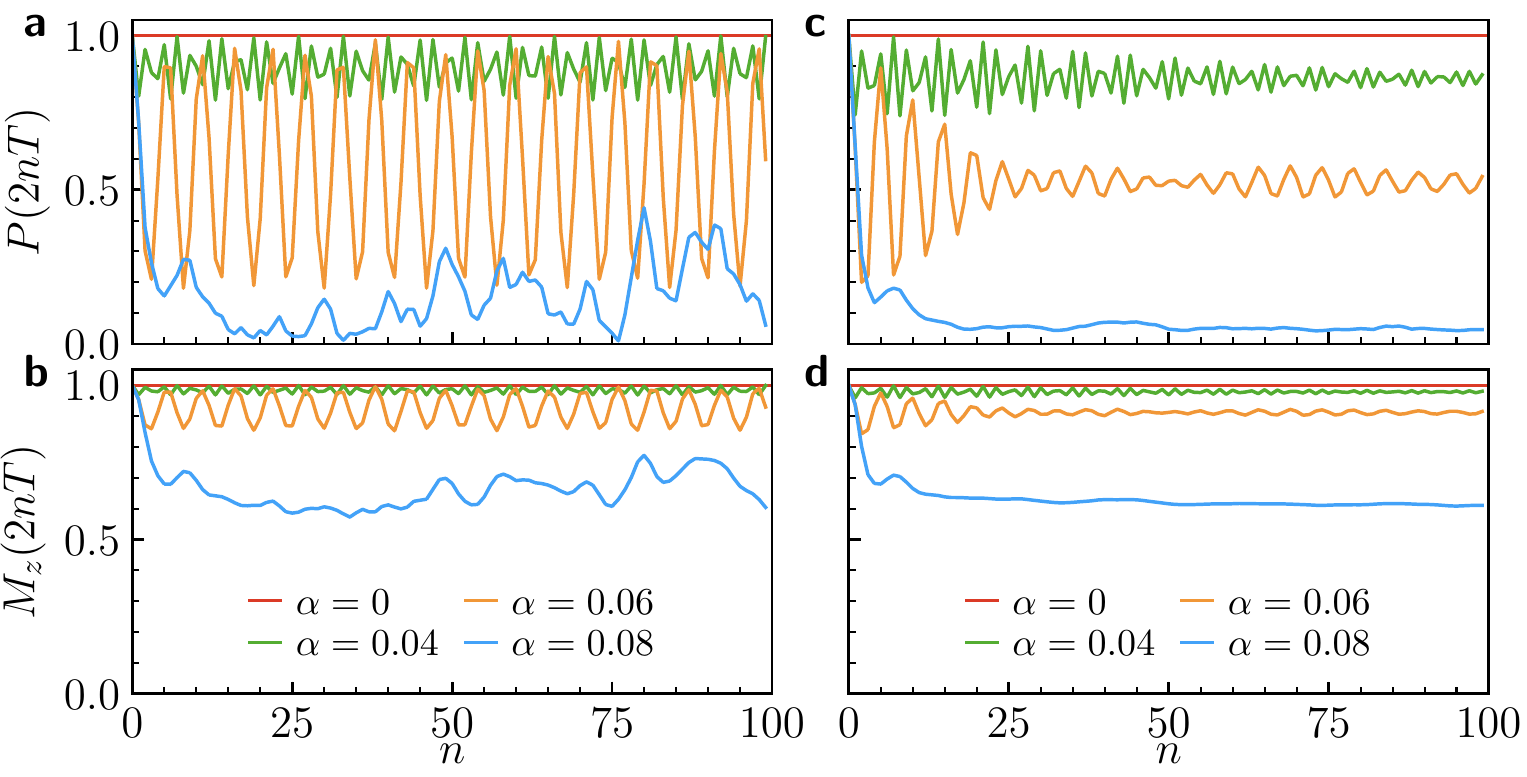}
  \caption{\label{fig4} Power-law potentials for $N=14$. (a,b) Uniform rotations 
  with $\bar{\theta}=0.95\pi$. $\alpha=0$ corresponds to the all-to-all interaction. With decreasing 
  $\alpha$, the results of the power-law potentials approach those of the all-to-all 
  interaction. All parameters are the same in (a) and (b). 
  (c,d) Inhomogeneous rotations with $\bar{\theta}=\pi$ 
  and $w_s=0.1\pi$.}
\end{figure}
 
Whereas we have been focusing on the all-to-all interaction, similar conclusions 
apply to a generic long-range interaction, provided that its range is much larger 
than the size of the system. For instance, with decreasing $\alpha$, the range of 
the a power-law potential in Eq.~(\ref{Hintp}) increases. When $\alpha=0$, it is 
equivalent to the all-to-all interaction. Fig.~\ref{fig4} shows the results 
for $N=14$. With decreasing $\alpha$ down to zero, $P(2nT)$ and $M_z(2nT)$ increase 
and eventually approach the result of the all-to-all interaction. A small $\alpha=0.04$ 
readily provides us with a good approximation of the all-to-all interaction in such 
a finite system. 

Both interactions and external drivings are crucial for DTCs. We hope that our 
work will stimulate more  studies of their interplays to access novel non-equilibrium 
quantum states with long coherent time.

We acknowledge C.-L.~Hung for helpful discussions on the precision of measuring frequencies. This work is supported by DOE DE-SC0019202, W.~M.~Keck Foundation, and the Center 
for Science of Information (CSoI), an NSF Science and Technology Center, under 
grant agreement CCF-0939370. C.~Lv acknowledges support from Purdue Research Foundation.

\clearpage
\pagebreak
\newpage
\widetext
\begin{center}
\textbf{\large 
Supplemental Material of ``An eternal discrete time crystal beating the Heisenberg limit''
}
\end{center}
\renewcommand{\theequation}{S\arabic{equation}}
\renewcommand{\thefigure}{S\arabic{figure}}
\renewcommand{\thetable}{S\arabic{table}}
\renewcommand{\theHequation}{Supplement.\theequation}
\renewcommand{\theHtable}{Supplement.\thetable}
\renewcommand{\theHfigure}{Supplement.\thefigure}
\setcounter{table}{0}
\setcounter{figure}{0}
\setcounter{equation}{0}

\onecolumngrid

\section{Onsite disorder }
\begin{figure}[b]
\includegraphics[width=\textwidth]{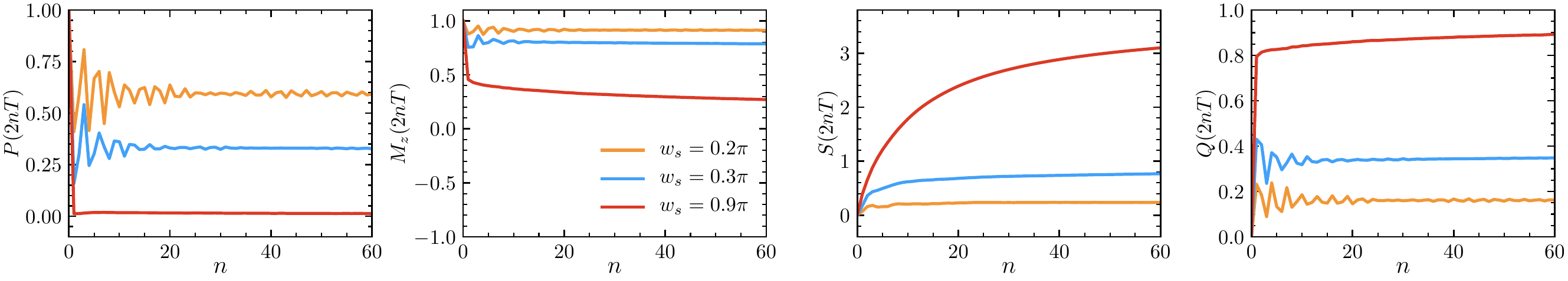}
\caption{\label{fig:disorder} Effects of inhomogeneous $\theta_i$ on MBL. 
$\theta_i$ is chosen from a uniform distribution $[-w_s+\bar{\theta}, \bar{\theta}+w_s]$. 
The onsite disorder is chosen from $[0, W]$, where $W=40J,$ $\alpha=3$, 
$\bar\theta=\pi$, $N=14$.}
\end{figure}

The onsite disorder is often considered in DTC to introduce many-body localization.
Since the coupling between l-bits decays exponentially with increasing their distance, 
this could slow down the thermalization, provided that $\theta_{i}$ is spatially uniform. 
However, this mechanism of suppressing the thermalization automatically weakens the 
synchronization between different spatial parts of the system. Thus, when $\theta_{i}$ 
has strong spatial inhomogeneities, the onsite disorder cannot stabilize the DTC. 
Consider the Hamiltonian, 
\begin{align}
H = 2J\sum_{i<j}\frac{S_i^zS_j^z}{|i-j|^\alpha} +\sum_n  \delta(t-nT)\sum_{i=1}^N \theta_{i}{S}^y_i + 2\sum_{i=1}^N\Delta_iS_{i}^z,
\end{align}
where  $\theta_{i}$ has a uniform distribution in $[\bar{\theta}-w_s, \bar{\theta}+w_s]$, 
similar to the main text. The onsite disorder, $\Delta_i$, has a uniform 
distribution in $[0,W]$.  As shown in Fig.~\ref{fig:disorder},  for a given $W$, 
with increasing $w_s$, $P(2nT)$ and $L_z(2nT)$ are suppressed down to zero. 
Meanwhile, the entropy $S(2nT)$ and $Q(2nT)$ grow faster, signifying the 
thermalization of the DTC.

\section{Scalings with particle numbers }

\subsection{Scalings at $t=2T$ }

\begin{figure} 
\includegraphics[width=\textwidth]{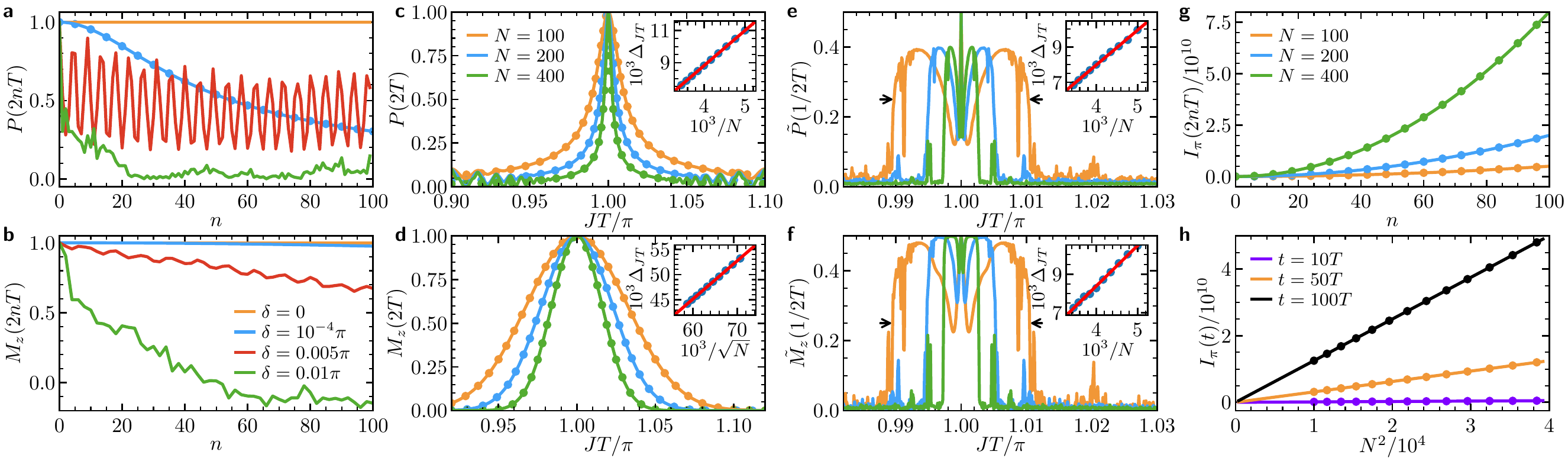}
\caption{\label{fig3:05} Sensitivity of the DTC to $JT$ when $\theta=\pi/2$.
Dots are analytical results and curves are the numerical results. 
(a,b) $P(2nT)$ and $M_z(2nT)$ as functions of $n$ for various $JT$.
When $|JT-\pi|\gg\pi/N$ (here N=200), both quantities quickly decrease down to zero. 
(c,d) $P(2T)$ and $M_z(2T)$ 
at a fixed time $t=2T$ as functions of $JT$. For a fixed $N$, both quantities 
are featured with narrow peaks centered at $JT=\pi$. Insets show the scalings 
of the widths $\Delta$ (full width at half maximum) of the peaks with $N$. (e,f)
The power spectra $\tilde{P}(1/2T)$ and $\tilde{M_z}(1/2T)$ are also featured with narrow peaks around $JT=\pi$. Whereas they exhibit non-monotonic behaviors near $JT=\pi$, both quantities vanish when 
$|JT-\pi|\gg\pi/N$. Insets show the scalings of the widths of the peaks with $N$. 
(g) The quantum Fisher information $I_{\pi}(2nT)$ as a function of $n$.
(h) $I_\pi(2nT)$ is proportional to $N^2$. $\theta_i=\pi/2$ is used in all panels.}
\end{figure}

We have analytically obtained how the dependence of $P(2T)$ ($M_z(2T)$) on $JT$ 
scales with the particle number $N$,
\begin{align}
 {P}'(\delta)&\equiv P(2T; JT=\pi+\delta)=(1+\frac{N^2\delta^2\sin^4\theta}{4})^{-1/2}e^{-\frac{1}{4N^2\delta^2\sin^4\theta+16}\sin^2(2\theta)\delta^2N^3}
 ,\label{p2t:theo}\\
 {M'_z}(\delta)&\equiv M_z(2T;JT=\pi+\delta)= \cos^2\theta+\sin^2\theta \cos(\cos\theta N\delta)e^{-\frac{N}{2}\delta^2\sin^2\theta}. \label{pz:theo}
\end{align}
When $\theta=\pi/2$, $\sin(2\theta)=0$. The exponential function in Eq.~(\ref{p2t:theo}) 
becomes identity, and ${P}'(\delta)=(1+\frac{N^2\delta^2\sin^4\theta}{4})^{-1/2}$. 
Thus, the peak width shown in Fig.~\ref{fig3:05} scales with $1/N$ . The same 
scaling applies to $\theta$ near $0$ and $\pi$. In contrast, when $\theta$ is away 
from 0, $\pi/2$ and $\pi$, the exponential function becomes dominant, and $P(2T)$ 
decays faster, as shown in Fig.~3(c) of the main text. In particular, the peak 
width of $ {P}(2T)$ in Fig.~3(e) scales with $1/N^{3/2}$.
 
In Eq.~(\ref{pz:theo}), the $N\delta$ term in the cosine function leads to a fast
oscillation, and the $N\delta^2$ term in the exponential function 
leads to the $1/N^{1/2}$ scaling of the profile of $M_z(2T)$, regardless of 
$\theta$, as shown in the insets of Fig.~\ref{fig3:05}(d) of this supplementary 
material and Fig.~3(d) of the main text.

To derive Eq.~(\ref{p2t:theo}) and Eq.~(\ref{pz:theo}), we consider an initial 
state, $\ket{\Psi(0^-)}=\prod_i\ket{\uparrow}_i=\ket{\frac{N}{2},\frac{N}{2}}$, 
where $L_z\ket{\frac{N}{2},l}=l\ket{\frac{N}{2},l}$ and $\frac{N}{2}=L$ is the 
total angular momentum.
\begin{align}
\begin{aligned}    
{P}'(\delta)=& |\bra{\Psi(0^-)}e^{-i(\pi+\delta)L_z^2}e^{-i\theta L_y}e^{-i(\pi+\delta) L_z^2}e^{-i\theta L_y}\ket{\Psi(0^-)}|^2\\
=&|e^{-i(\pi+\delta)(\frac{N}{2})^2}\bra{\Psi(0^-)}e^{-i\theta L_y}e^{-i(\pi+\delta) L_z^2}e^{-i\theta L_y}\ket{\Psi(0^-)}|^2.
\end{aligned}
\end{align}

As discussed in the main text, 
$e^{-i\pi L_z^2}=e^{-i\pi L_z}$ is satisfied for any even particle number $N$. 
When $\delta$ is small, $e^{-i\delta L_z^2}$ can be written as
$ e^{-i\delta L_z^2}\approx\int_{-N\delta}^{N\delta} dke^{ik^2/4\delta}\frac{1}{2\pi}\frac{\sqrt{\pi}}{\sqrt{i\delta}}e^{-ikL_z}$.
We thus obtain
\begin{align}\label{operator:1}
P'(\delta) &\approx|e^{-i(\pi+\delta)(\frac{N}{2})^2}\int_{-N\delta}^{N\delta} dke^{ik^2/4\delta}\frac{1}{2\pi}\frac{\sqrt{\pi}}{\sqrt{i\delta}}I(\theta, k)|^2,\\  
I(\theta, k) &=\bra{\Psi(0^-)}e^{-i\theta L_y}e^{-i\pi L_z}e^{-i k L_z}e^{-i\theta L_y}\ket{\Psi(0^-)}.
\end{align}
 Note that
$e^{-i\theta L_y}\ket{\Psi(0^-)}=\ket{\theta, 0}_c$, $ e^{i\pi L_z} e^{i\theta L_y}\ket{\Psi(0^-)} =(-1)^{\frac{N}{2}} \ket{\theta, 0}_c$,
we obtain
\begin{equation}\label{weight:1}
 I(\theta,k) = (-1)^{\frac{N}{2}}\bra{\theta,0}_ce^{-i k L_z}\ket{\theta,0}_c=(-1)^{\frac{N}{2}}e^{i k \frac{N}{2}}(\frac{1}{1+\alpha})^{N}(e^{-i k}\alpha+1)^{N},
\end{equation}
where $\alpha\equiv \tan^2\frac{\theta}{2}$ and
\begin{align}
\ket{\theta, \phi}_c = \sum_{l=-N/2}^{N/2}\sqrt{\frac{N!}{(\frac{N}{2}+l)!(\frac{N}{2}-l)!}} (\cos\theta)^{\frac{N}{2}+l}(\sin\theta)^{\frac{N}{2}-l}e^{i\phi(\frac{N}{2}-l)}\ket{l}
\end{align}
is a coherent state pointing along $\theta, \phi$. In the large $N$ limit, 
\begin{equation}\label{weight:result}
    I(\theta,k) \approx (-1)^{\frac{N}{2}}e^{-i k \frac{N}{2}+i\frac{N}{1+\alpha}k}e^{-\frac{N}{2}\frac{\alpha}{(1+\alpha)^2}k^2}, 
\end{equation}
which represents a narrow Gaussian centered at $k=0$.  Substituting $I(\theta,k)$ 
in Eq.~(\ref{operator:1}) by Eq.~(\ref{weight:result}), we obtain Eq.~(\ref{p2t:theo}).

As for ${M'_z}(\delta)$, 
using the time evolution operator $U(T)=e^{-iL_z^2JT}e^{-iL_y\theta}$, we obtain 
the Heisenberg equations, which provide us with the nonlinear 
recursion relations as shown in \cite{Haake},
\begin{equation}\label{eom:0}
\begin{split}
    & L_x'=U^{-1}(T)L_xU(T)=\frac{1}{2}(L_x \cos\theta+L_z \sin\theta+iL_y)e^{i2JT(L_z\cos\theta-L_x\sin\theta+\frac{1}{2})}+h.c.\\
    & L_y'=U^{-1}(T)L_yU(T)=\frac{1}{2i}(L_x \cos\theta+L_z \sin\theta+iL_y)e^{i2JT(L_z\cos\theta-L_x\sin\theta+\frac{1}{2})}+h.c. \\
    & L_z'=U^{-1}(T)L_zU(T)=L_z\cos\theta-L_x\sin\theta. \\
\end{split}
\end{equation}
Since $M_z(2T)=\frac{2}{N}\bra{\Psi(0^-)}U^{-1}(T)U^{-1}(T)L_zU(T)U(T)\ket{\Psi(0^-)}$, we obtain, 
\begin{equation}\label{Lz2t:0}
    \begin{split}
    &\begin{split}
        M'_z(\delta)=\frac{2}{N}\bra{\Psi(0^-)}&[ (L_z\cos\theta-L_x\sin\theta)\cos\theta\\
        &-(\frac{1}{2}(L_x \cos\theta+L_z \sin\theta+iL_y)e^{i2JT(L_z\cos\theta-L_x\sin\theta+\frac{1}{2})}+h.c.)\sin\theta ]\ket{\Psi(0^-)}
    \end{split}\\
    &=\cos^2\theta-(-1)^{\frac{N}{2}}\sin\theta[\sin\theta \cos^{N}\frac{\Theta}{2}\cos(JT-N\Phi)+(\cos\theta+1)\cos^{N}\frac{\Theta}{2}\tan\frac{\Theta}{2}\cos(\Phi+JT-N\Phi)]\\
    &=\cos^2\theta+(-1)^{\frac{N}{2}}\sin\theta \cos^{N}(\frac{\Theta}{2})[\sin\theta \cos(\delta-N\Phi)+(\cos\theta+1)\tan\frac{\Theta}{2}\cos(\Phi+\delta-N\Phi)]   \\
    & \approx \cos^2\theta+\sin^2\theta \cos(\cos\theta N\delta)e^{-\frac{N}{2}\delta^2\sin^2\theta}
    \end{split}
\end{equation}
where $\Theta=\arccos(\cos^2\theta+\cos(2\delta)\sin^2\theta)$, $\Phi=
\arctan(\frac{-\sin\theta \sin(2\delta)}{\cos\theta \sin\theta(-1+\cos(2\delta))})$.
The expression which contains $\Theta$ and $\Phi$ is exact for any $\theta$ and $\delta$'s. The final approximation comes from $\Theta^2=4\sin^2\theta \delta^2+O(\delta^4)$, $\Phi=\pi/2-\cos\theta\delta+O(\delta^3)$ and $\cos^{N}\frac{\Theta}{2}\approx e^{-N\delta^2\sin^2\theta/2 }$ when $\delta$ is small and $N$ is large. The overall 
profile as shown in Fig.~3(d) of the main text is thus given by $e^{-\frac{N}{2}\sin^2\theta \delta^2}\sin^2\theta+\cos^2\theta$.

\subsection{Scalings of $\tilde{P}(\frac{1}{2T})$}
\begin{figure} 
\includegraphics[width=\textwidth]{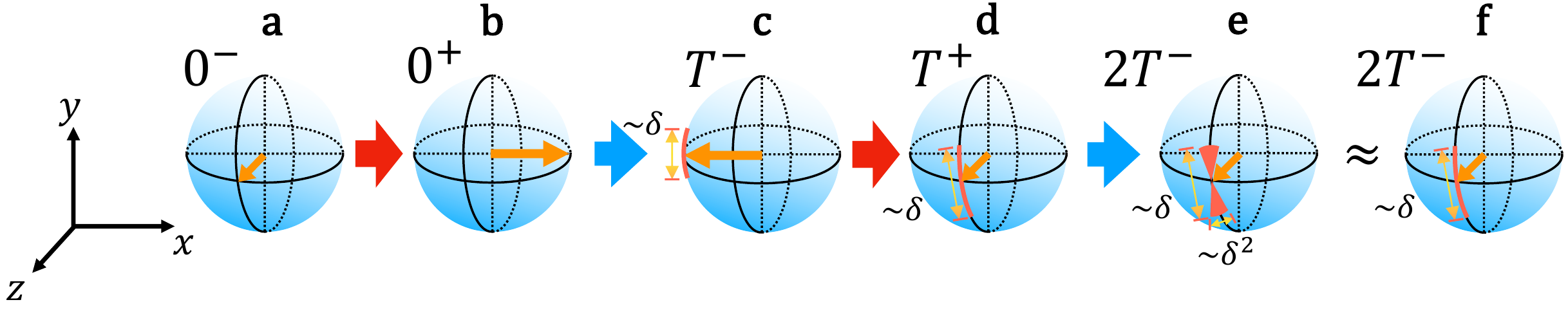}
\caption{\label{fig:approx} The approximation used to derive $P(2nT)$. When $t=T^-$,
 the nonlinear operator $e^{-iJTL_z^2}$ creates a superposition of coherent states, 
 which spans a length scale $\sim \delta$ in the latitude direction, as shown in (c). 
 This length scale is transferred to the longitude direction around the north pole 
 by the pulse at $t=T^+$, as shown in (d). Then the nonlinear operator creates a 
 superposition of coherent states in the region highlighted by the red color at 
 $t=2T^-$. The length scales of this region in the longitude and latitude 
 directions are $\delta$ and $\delta^2$ respectively, as shown in (e). Replacing 
 the second nonlinear operator $e^{-iJTL_z^2}$ by $e^{-i\pi L_z^2}$, as shown in 
 (f), we have ignored the expansion of the wavefunction in the latitude direction 
 that gives rise to a high order correction to $P(2nT)$ at small times.}
\end{figure}
We have also obtained an analytical form for $\tilde{P}(f)$, the Fourier transform of $P(2nT)$. 
As shown in Fig.~\ref{fig:approx}, starting from an initial state at the north pole, the state at $t=2T^-$ covers a finite small region near the north pole, if $\delta=JT-\pi$ is small. The length scales of the longitude and latitude directions are proportional to $\delta$ and $\delta^2$, the latter of which can be ignored in the small $\delta$ limit. Thus, we make use of the following approximation to capture the dynamics in the small $\delta$ limit, 
\begin{equation}\label{p2nt:approx}
    U(2T)=e^{-iJTL_z^2}e^{-i\theta L_y}e^{-iJT L_z^2}e^{-i\theta L_y}\approx e^{-i\pi L_z^2}e^{-i\theta L_y}e^{-iJT L_z^2}e^{-i\theta L_y}.
\end{equation}
$P(2nT)$ is written as
\begin{equation}\label{p2nt:0}
    P(2nT)=|\bra{\Psi(0^-)}(e^{-i\pi L_z^2}e^{-i\theta L_y}e^{-i(\pi+\delta) L_z^2}e^{-i\theta L_y})^n\ket{\Psi(0^-)}|^2.
\end{equation}
Using the identities, $e^{-i\pi L_z^2}=e^{-i\pi L_z}$ and $e^{-i\pi L_z}e^{-i\theta L_y}e^{-i\pi L_z}=e^{i\theta L_y}$, the equation above can be written as 
\begin{equation}\label{p2nt:1}
    P(2nT)=|\bra{\Psi(0^-)}(e^{i\theta L_y} e^{-i \delta L_z^2}e^{-i\theta L_y})^n\ket{\Psi(0^-)}|^2=|\bra{\Psi(0^-)}e^{i\theta L_y} e^{-i n\delta L_z^2}e^{-i\theta L_y}\ket{\Psi(0^-)}|^2.
\end{equation}
Applying Eq.~(\ref{p2t:theo}), we obtain 
\begin{equation}\label{p2nt:theo}
    P(2nT)=e^{-\frac{n^2\sin^2(2\theta)\delta^2N^3}{4N^2n^2\delta^2\sin^4\theta+16}}(1+\frac{n^2N^2\delta^2\sin^4\theta}{4})^{-1/2}.
\end{equation}
Eq.~(\ref{p2nt:theo}) recovers Eq.~(\ref{p2t:theo}) when $n=1$.  As shown in Fig.~\ref{fig3:05}(a), this expression well captures the initial decay of $P(2nT)$. However, it cannot describe the revival of $P(2nT)$ in later times for certain $JT$ due to the made approximation in Eq.~(\ref{p2nt:approx}).  

The power spectrum is therefore written as 
\begin{equation}\label{p2nt:peakheight}
    \tilde{P}(1/2T)=\frac{1}{M}\sum_{n=0}^{M-1}P(nT)e^{i\frac{2\pi}{2T}nT}=\frac{1}{M}\sum_{n=0}^{M-1}P(nT)(-1)^n\approx \frac{1}{M}\sum_{n=0}^{M/2-1}P(2nT),
\end{equation}
where $M$ is the cutoff required in numerics.
In the last step, we have used the fact that, for small $n$, $\ket{\psi(2nT+T)}$ is located at a place on the Bloch sphere away from the north pole, provided that $\theta$ is not small, and thus, 
$P(2nT+T)=|\bra{\Psi(0^-)}\ket{\psi(2nT+T)}|^2\approx 0$.

When $nN\delta\ll 1$ and $\theta\neq 0, \pi/2, \pi$,  Eq.~(\ref{p2nt:theo}) becomes 
$P(2nT)=e^{-\frac{n^2\sin^2(2\theta)\delta^2N^3}{16}}$, 
and Eq.~(\ref{p2nt:peakheight}) is rewritten as
\begin{equation}\label{p2nt:phtheo_general}
    \tilde{P}(1/2T)\approx \frac{2\sqrt{\pi}\text{Erf}(\frac{1}{8}\delta M N^{3/2}\sin (2\theta))}{\sin(2\theta)\delta M N^{3/2}},
\end{equation}
In the limit $M\rightarrow \infty$, $\tilde{P}(1/2T)\rightarrow \frac{1}{2\sqrt{\pi}}e^{-\frac{1}{64}\delta^2N^3M^2}$. To compare with numerical result, we choose $\theta = \pi/4$ and $M=200$. Eq.~(\ref{p2nt:phtheo_general}) becomes
\begin{equation}\label{p2nt:phtheo}
    \tilde{P}(1/2T)\approx \frac{2\sqrt{\pi}\text{Erf}(25\delta N^{3/2})}{200\delta N^{3/2}},
\end{equation}
which shows the $1/N^{3/2}$ scaling. $\text{Erf}$ is the error function. The comparison between this analytical result and the numerical one is shown in Fig.~3(e) of the main text.

\begin{figure}[h]
\includegraphics[width=\textwidth]{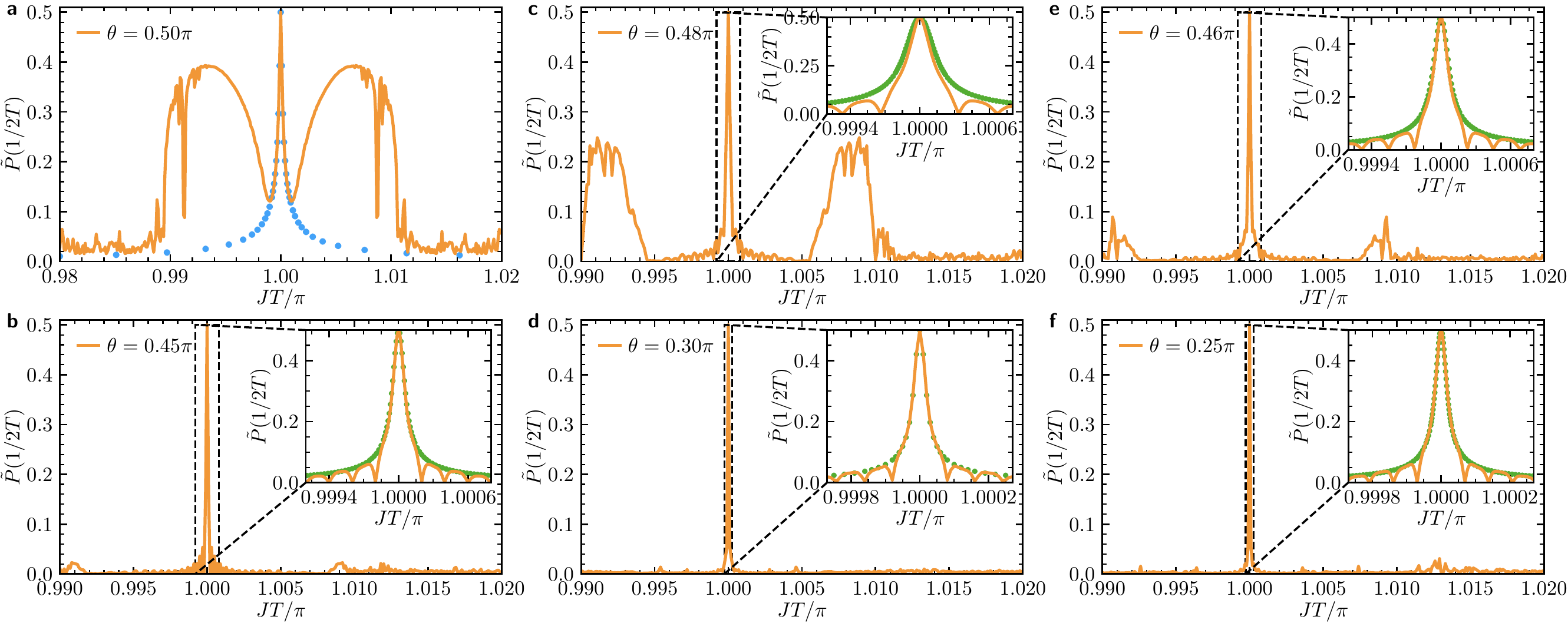}
\caption{\label{fig:peakheight} The power spectra $\tilde{P}(1/2T)$ for different 
pulses for $N=100$, $M=200$. (a) It is identical to the curve for $N=100$ in 
Fig.~\ref{fig3:05}(e). The central sharp peak at $JT=\pi$ for $\pi/2$ pulse is well 
approximated by the analytical result (blue dots) shown in Eq.~(\ref{p2nt:phtheo_pi2}). 
(b-f) When $\theta$ decreases, the two broader peaks gradually vanish. 
When the exponential term in Eq.~(\ref{p2nt:theo}) dominates, the central peak is 
described by Eq.~(\ref{p2nt:phtheo_general}) (green dots). 
Enlarging (f) around $JT=\pi$ gives rise to Fig.~3(e) of the main text.}
\end{figure}
  
When $\theta=\pi/2$, the exponential term in Eq.~(\ref{p2nt:theo}) becomes identity.
We obtain 
\begin{equation}\label{p2nt:phtheo_pi2}
    \tilde{P}(1/2T)\approx \frac{\text{arcsinh}(\delta M N/4)}{\delta M N/2}.
\end{equation}

As mentioned in Fig.~\ref{fig3:05}, when $\theta=\pi/2$, the dependence of  $\tilde{P}
(1/2T)$ on $JT$ is not monotonic. With increasing $\delta$, $\tilde{P}
(1/2T)$ first quickly decreases and then increases before it eventually vanishes when $\delta>\pi/N$. Eq.~(\ref{p2nt:phtheo_pi2}) captures the narrow peak, whose width is much smaller than $\pi/N$, near $\delta=0$. The broader peak scales with $
1/N$ as shown in Fig.~\ref{fig3:05}(f) of the main text. 
When $\theta$ deviates from
$\pi/2$, the broader peak gets suppressed as shown in Fig.~\ref{fig:peakheight}. 
When $\theta=\pi/4$, only the central narrow peak is visible, whose width scales with 
$1/N^{3/2}$, as discussed before. 

$M_z(2nT)$ and $\tilde{M}_z(1/2T)$ do not have simple analytical forms. We have numerically evaluated them and the scaling of $\tilde{M}_z(1/2T)$ with $N$ is shown in Fig.~\ref{fig3:05}(b, e).

\section{quantum Fisher information} 
When $JT=\pi$, the quantum Fisher information is written as 
\begin{align}
I_{\pi}(2nT)&=\lim_{\epsilon\rightarrow 0}4\frac{1-F_{\epsilon}}{\epsilon^2},\\
F_{\epsilon}&=|\bra{\Psi(0^-)}U_{\pi}(-2nT)U_{\pi+\epsilon}(2nT)\ket{\Psi(0^-)}|^2. 
\end{align}
As $U_{\pi}(2nT)\ket{\Psi(0^-)}=\ket{\Psi(0^-)}$ or equivalently, $\bra{\Psi(0^-)}U_{\pi}(-2nT)=\bra{\Psi(0^-)}$, the Loschmidt echo is identical to the quantum memory of the initial state,  
$F_{\epsilon} = |\bra{\Psi(0^-)}U_{\pi+\epsilon}(2nT)\ket{\Psi(0^-)}|^2$. Using Eq.~(\ref{p2nt:theo}) and replacing $\epsilon$ by $\delta$, we obtain
\begin{align}
I_{\pi}(2nT) = \lim_{\delta\rightarrow0}4\frac{1-P(2nT)}{\delta^2} = \frac{n^2N^3\sin^2(2\theta)}{4} + \frac{n^2N^2\sin^4\theta}{2}.
\end{align}

\end{document}